# Multiobjective Based LQR Design for Fractional Order Systems with Perturbations using PESA-II Algorithm

**Lalitesh Kumar[1], Prawendra Kumar[2]**

[1] Zhejiang University, 38 Zheda Road, Hangzhou, 310058, China (E-mail: lalit2281985@hotmail.com, 11932075@zju.edu.cn).
[2] Politechnica Gdanska, Gdansk, 88-266, Poland, (E-mail: prawendrak7@gmail.com).

**Abstract**— In this work, perturbed LQR design for fractional-order systems with multiple performance indices is proposed. The first one performance index constitutes a weighted sum of ITAE (Integral Time of Absolute Error) and ISCO (Integral Squared Controller Output), chosen to minimize the large oscillations as well as control efforts. The second one ensures robust stability with the minimization of singular values of the return ratio matrix at the plant's input. The third and the last performance index aims minimization of maximum singular values of the perturbation transfer function at the plant's output, so as to guarantee no closed-loop right half plane zeros. The multi-objective optimization problem with these three objective functions is framed in both frequency as well as time domain, and an optimal solution is proposed via Pareto Envelope Based Selection Algorithm-II (PESA-II). The simulation results validate the performance of the proposed algorithm in terms of perturbations and are compared with those with the general cases of the weighting matrices.

## 1. INTRODUCTION

Since the beginning of the twenty-first century, the fractional-order control systems have remained the focus of research particularly aimed at fractional re-design of PID controllers. The resurge is evident from numerous papers published in the area of design of fractional order controllers in recent years with emphasis over the fractional order PID (FO-PID) controllers. The design of fractional order controllers is relatively simple with improved system performance. I. Podlubny [1] proposed the design concept for the FO-PID controller, which is widely for the design of FO-PID controllers [1]. The LQ regulators achieve optimal desired state dynamics that approach zero under minimal control effort by proper selection of weighting matrices. For fractional order systems (FOSs), state-space models have been derived from non-integer order transfer functions. Li and Chen [2] discussed the design of LQR for FOS via computation of fractional Euler-Lagrange relations. The analytical asymptotic solutions for right-sided fractional equations were derived with discussions over error considerations. Kiani et al. [3] proposed the design of linear quadratic Gaussian (LQG) for FOSs via introducing a fractional differential in control law. However, their design considered only linear system dynamics and didn't admit guaranteed minimum energy consumption.

Based on the minimization of two conflicting performance indices over time domain with the help of multi-objective optimization, weighting matrices have been optimized for the extremum selection to design an LQR controller-based fractional order PID controller [4]. Their design preserved the optimality of state dynamics of LQR along-with optimally tuned fractional order PID controller. Interestingly it was shown that design tradeoff exists in two performance indices as an improvement in one result in degradation in other. However, robustness issues were not addressed explicitly. Owing to excellent captivating robustness properties of LQRs, scads of researchers have shown interests in investigating the robustness issues in LQ regulators with deliberate perturbations given to their state-space formulation [5]-[8]. Further, several improvements were suggested thereafter [9]-[10] instability robustness margins of LQR in systems incorporated with perturbations. Research efforts have been reported in literature towards the merging of optimal PID tuning with LQ regulator design, but most of them were formulated in the time domain only. This paper addresses the LQR design for fractional order systems as a part of multi-objective optimization in both the time domain as well as frequency domain. Fractional order sliding mode control based upon the LQR for the linearized model of a class of uncertain nonlinear systems were proposed to guarantee quick convergence with robustness [11]. Further, analytical stability results were provided through the Lyapunov approach and the resulting sliding mode controller efficiently rejected system uncertainties and reduced control chattering. LQR for the fractional-order system with a certain time delay at plant input was proposed without any approximation and estimation [12]. The Design of cascaded FO PID-IO PID was formulated as a multi-objective problem for minimization of control effort and ISE to achieve robust stability [13]. Discrete PID based on the LQR with fractional order integral performance index and real coded genetic algorithm (GA) optimized weighting matrices [14].

In the present paper, the fractional-order system is converted into a commensurate order state-space model for use in subsequent design and optimization formulation. The design of the perturbed LQR system and robust stability have been analyzed along with performance. The perturbations tolerated by LQR are highly related to diagonal elements in the weighting matrices used in the quadratic performance index. The remaining part of the paper is organized as follows: Section 2 focuses on the LQR concept for fractional-order systems and the PESA-II algorithm, section 3 deals with multi-objective problem formulation followed by case studies and discussion in section 4. Section 5 discusses the conclusion and future directions.

## 2. LQR FOR FRACTIONAL ORDER SYSTEMS

The LQRs have gained significant acceptance in modern control theory owing to their guaranteed stability margins, meaningful representation of a performance in terms of entries in Q and R matrices and above all linear control relations. The frequency-domain interpretation or robustness margin is conceptualized in terms of return difference relations, which analyze important feedback properties. In systems incorporating LQR, asymptotic behavior is observed as weighting functions approach threshold values, and thus demands the selection of values to achieve better performance [5]. Generally, the problem is formulated using a quadratic performance index and may admit bilinear terms. LQR optimizes the performance index (characterized in terms of Q and R matrices) through the minimal deviation of state trajectories from the nominal operating point. However, the mapping between standard time/frequency domain performance requirements and entries of Q and R matrix is not direct and thus the closed-loop response may not be good for every choice of weighting matrix. Improper selection of Q and R may cause high overshoot and oscillations. Thus, it becomes obvious to choose weighting matrices so as to suit time domain specifications. Hit and trial or multi-objective formulations may be required to tune these matrices to admit excellent tracking even for setpoint changes. Population-based heuristic techniques such as Genetic Algorithms (GAs) have been used for optimization and selection of optimum weighting matrices for a class of LQR formulations for the design of fractional order PID controller [15][16]. Li et al. [17] proposed the multi-objective evolutionary algorithm (MOEA) based approach for the selection of weighting matrices in LQR with application to benchmark problem. GAs based selection of optimum weighting matrices of LQR were proposed for a wide variety of practical applications such as controlling Buck converter and collection of earthquake data [18][19] respectively. The performance index for an LQR for searching optimal control energy alongside state deviations terms is as [7]:

$$J = \int_0^{t_f} [x^T(t)Qx(t) + u^T(t)Ru(t)]dt \qquad (1)$$

Here, Q is positive semi-definite and R is positive definite matrices. The first term in the above expression represents the total energy of the system as it evolves over time horizon, whereas the second term represents the amount of control input energy spent over the time horizon $0 \leq t \leq t_f$. For integer order state – space systems the performance index (1) is valid and for fractional order state-space systems the control law is selected such that it can be converted into commensurate order state-space autonomous systems.

Let a fractional order system is represented as following $T(s) = \frac{Y(s)}{U(s)} = \frac{p_0 + p_1 s^{m_1} + \ldots\ldots p_M s^{m_M}}{q_0 + q_1 s^{m_1} + \ldots\ldots + q_{N-1} s^{m_{N-1}} + s^{m_N}} = \frac{p(s)}{q(s)}$ The fractional orders $m_1, m_2 \ldots\ldots m_N$ are positive real numbers called as explicit orders. Implicit orders can be defined as following:

$$n_1 = m_1 \ldots\ldots n_i = m_{i-1} \ldots\ldots n_N = m_{N-1}$$

The fractional order differential equation is incomplete without defining the fractional order integral operator. Realization of integral operator could be in time domain or in frequency domain. In this paper time domain realization has been considered for fractional order integral operator as defined in [20], [21].

$$I^n(s) = 1/s^n, \quad 0 < n < 1 \qquad (2)$$

Now pseudo state-space model of the fractional order systems can be written as follows:

$$x_1(t) = x(t), \quad x_2(t) = D^{n_1}(x_1(t))$$
.
.
$$x_i(t) = D^{n_{i-1}}(x_{i-1}(t))$$
.
.
$$x_N(t) = D^{n_{N-1}}(x_{N-1}(t))$$

$$D^{n_N}(x_N(t)) = -q_0 x_1(t) \ldots\ldots - q_{N-1} x_N(t) + u(t) \quad \text{and} \qquad (3)$$

$$x_1(t) = I^{n_1}(x_2(t))$$
$$\vdots$$
$$x_{i-1}(t) = I^{n_{i-1}}(x_i(t)) \tag{4}$$
$$\vdots$$
$$x_{N-1}(t) = I^{n_{N-1}}(x_N(t))$$
$$x_N(t) = I^{n_N}(u(t))$$

Now pseudo state-space equation is as given following:
$$D^n(X(t)) = AX(t) + Bu(t) \text{ with} \tag{5}$$

$$X(t) = \begin{bmatrix} x_1(t) \\ \vdots \\ x_i(t) \\ \vdots \\ x_N(t) \end{bmatrix} \text{ and } D^n(X(t)) = \begin{bmatrix} D^{n_1}(x_1(t)) \\ \vdots \\ D^{n_i}(x_i(t)) \\ \vdots \\ D^{n_N}(x_N(t)) \end{bmatrix}$$

$$A = \begin{bmatrix} 0 & 1 & 0 & \cdots & 0 \\ 0 & 0 & 1 & \cdots & 0 \\ \vdots & & & & \vdots \\ 0 & 0 & & \cdots & 1 \\ -q_0 & -q_1 & & \cdots & -q_N \end{bmatrix}, \quad B = \begin{bmatrix} 0 \\ 0 \\ \vdots \\ 0 \\ 1 \end{bmatrix}$$

The output is $y(t) = C^T X(t)$ with
$$C^T = [p_0 \ldots\ldots p_M \ 0 \ldots 0]$$

The following examples from [21] [22] is taken to prove effectiveness of the present design method. The first example is a fractional order model of a typical system is given by

$$G_{FO}(s) = \frac{1}{2012.409s^{1.8063} + 107.2882s^{0.93529} + 1.0305} \tag{6}$$

The commensurate order state space equation of the above system without approximation is given as below:

$$\begin{bmatrix} D^{0.93529} x_1(t) \\ D^{0.87101} x_2(t) \\ D^{0.93529} x_3(t) \end{bmatrix} = \begin{bmatrix} 0 & 1 & 0 \\ 0 & 0 & 1 \\ -0.0005121 & -0.05331 & 0 \end{bmatrix} \begin{bmatrix} x_1(t) \\ x_2(t) \\ x_3(t) \end{bmatrix} + \begin{bmatrix} 0 \\ 0 \\ 1/107.2882 \end{bmatrix} u(t) \quad y(t) = \begin{bmatrix} 1 & 0 & 0 \end{bmatrix} \begin{bmatrix} x_1(t) \\ x_2(t) \\ x_3(t) \end{bmatrix} \tag{7}$$

Another example is taken as following in the equation (8) and the commensurate order state-space equation of 6th order (q=0.32) is as following [23]:

$$G_{FO}(s) = \frac{s^{0.32} + 5}{100s^{1.92} + 20s^{0.96} - 5s^{0.64} + 1} \tag{8}$$

The state-space model is as follows

$$D^q x(t) = \begin{bmatrix} 0 & 0 & -0.2 & 0.05 & 0 & -0.01 \\ 1 & 0 & 0 & 0 & 0 & 0 \\ 0 & 1 & 0 & 0 & 0 & 0 \\ 0 & 0 & 1 & 0 & 0 & 0 \\ 0 & 0 & 0 & 1 & 0 & 0 \\ 0 & 0 & 0 & 0 & 1 & 0 \end{bmatrix} x(t) + \begin{bmatrix} 1 \\ 0 \\ 0 \\ 0 \\ 0 \\ 0 \end{bmatrix} u(t) \tag{9}$$

$$y(t) = \begin{bmatrix} 0 & 0 & 0 & 0 & 0.01 & 0.05 \end{bmatrix} x(t)$$

Now fractional order control law and matrix Riccatti equation can be represented as given in the equation 10.

$$M^T U + UM + -UBR^{-1}B^T U + Q = 0$$
$$u(t) = -R^{-1}B^T U x(t) + K_{FO}\frac{d^\alpha x(t)}{dt^\alpha} \qquad (10)$$

Now to find a better control law we have to make the selection of weighting matrices *Q* and *R* by giving proper weightings. However, the control law determined from the above equation is fully reliant on the equation (1) but control design engineer can emphasize on the attenuation of the responses to the structural property and that is of substantial importance. In the present LQR design approach the design variables are chosen as diagonal elements of matrix *Q* and weighting factor matrix *R*. To solve the Riccatti equation we have to assign proper weightings to the matrices *Q* and *R*. The P matrix is been calculated in the MATLAB.

Now let us consider the Q and R matrices as follows:

$$Q = \begin{bmatrix} Q_1 & 0 & 0 & 0 & 0 & 0 \\ 0 & Q_2 & 0 & 0 & 0 & 0 \\ 0 & 0 & . & 0 & 0 & 0 \\ 0 & 0 & 0 & . & 0 & 0 \\ 0 & 0 & 0 & 0 & . & 0 \\ 0 & 0 & 0 & 0 & 0 & Q_n \end{bmatrix} \text{ and } R = \begin{bmatrix} R_1 & 0 & 0 & 0 \\ 0 & R_2 & 0 & . \\ . & . & . & . \\ 0 & 0 & . & R_n \end{bmatrix}$$

### *1.1 Fractional Order Systems*

It is a generalization of differential and integral fractional order and $a$ and t are lower and upper terminals of the operations respectively. The integro-differential will be represented as given in equation 11. Where in general $\alpha \in \mathfrak{R}$, however, it could be imaginary also. In general, there are two definitions of fractional order differetio-integral calculus is in use widely.

$$_aD_t^\alpha = \begin{cases} \dfrac{d^\alpha}{dt^\alpha} & \mathfrak{R}(\alpha) > 1 \\ 1 & \mathfrak{R}(\alpha) = 0 \\ \int_a^t (dt)^{-\alpha} & \mathfrak{R}(\alpha) < 0 \end{cases} \qquad (11)$$

The Grunwald-Letnikov definition and Riemann-Liouville definition [1], is given below,

$$_aD_t^\alpha = \lim_{h \to 0} h^{-\alpha} \sum_{i=0}^{\left[\frac{t-a}{h}\right]} (-1)^i \binom{\alpha}{i} f(t-hi) \qquad (12)$$

where [.] represents integer value and

$$_aD_t^\alpha = \frac{1}{\Gamma(n-\alpha)} \frac{d^n}{dt^n} \int_a^t \frac{f(\tau)}{(t-\tau)^{\alpha-n+1}} d\tau \qquad (13)$$

For $(n-1 < \alpha < n)$ and $\Gamma(.)$ is the gamma function. The Laplace transformation is used routinely to solve the integral and differential equations of engineering problems.

### *1.2 PESA-II Multi-objective Optimization*

Pareto Envelope Based Selection Algorithm-II (PESA-II) is an evolutionary optimization algorithm which solves multi-objective based optimization problems, based on the concept of genetics, and together with Pareto envelope-based selection. This algorithm is a modified version of PESA in which region-based selection has been used for assigning an individual selective fitness. The region of objective space is called hyperbox to which selective fitness is to be assigned. This type of selection ensures the spread of development along Pareto frontier in a better manner than individual selection. An archive consisting of non-dominated solutions is being maintained and selection will be done from this archive only. The selective fitness which is also called as 'squeeze factor' of an individual i.e. other solutions will capture the same hyper box. The tournament selection can be done by making this squeeze factor small.

After the selection of a region (or hyper cubes) is over now it is time to select the resulting individual from archives to perform genetic operations and the selection is being done randomly. For the selection of individual, it is generally taking care of to select a most isolated individual as compared to a highly crowded one and for this tournament, selection will be done by binary methods. Along with the standard tasks like crossover and mutation rates this algorithm has two measures pertaining to population size and one measure pertaining to the crowding strategy of the hyper-grid. Description of the algorithm is given in the following steps:

1. The first step is the initialization of the external population and evaluation of the generated internal population with $C_I$ chromosomes.
2. This step involves the incorporation of the initial population (non-dominated members from the archive) into the external population.
3. If reached to a termination point, then stop, return to the chromosomes in the external population. Otherwise, erase the existing contents of the internal population, and reiterate the step 4 until the generation of the new solution $C_I$.
4. Select new parents from the external population and repeat the crossover and mutation step to produce a new child string. Set the probability of this process as *k* and make the selection of one parent with a new probability *1-k*, perform the mutation of this parent to produce a child string.
5. Return to step 2.

Now after the above steps, a candidate solution may permit to enter the archive only if the candidate has non-dominancy over the internal population. Any dominated candidate will be removed from the archive.

### 3. PROBLEM FORMULATION

The perturbed LQR system with perturbation transfer function *L(s)* is shown in figure 1. The new lower bound of singular values for return difference matrix has been proved for perturbed LQR systems to guarantee the stability margins [5][6]. H. K. Khalil made a question mark on the robust stability of the LQR systems modelling errors [7]. To analyze robust stability of LQR systems it is necessary to study the perturbations effect on the algebraic Riccatti equation when both the weighting matrices are concerned with the design procedure. It is anticipated that the singular value concept of return difference method is very useful to alleviate the robustness issue if any developed due to modelling errors and some realistic perturbations. In general, LQR systems possess high robust stability and have optimal control effort and also it minimizes state trajectories in optimal fashion but it does not achieve the acceptable time domain specifications always and might constitutes high overshoot and oscillations. Hence, we need to choose weighting matrices in a fashion to satisfy some additional optimization performance indexes in terms of time domain specifications. One such performance criteria can be defined as weighted sum of integral time absolute error (ITAE) and integral of squared controller output (ISCO) as formulated in the following equation.

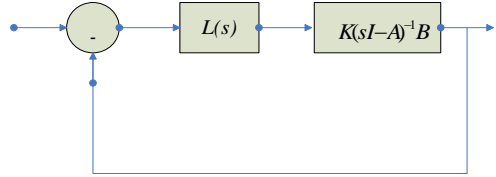

Figure 1. LQR system with perturbation

$$J_1 = \int_0^\infty \left[ S_1 * t|e(t)| + S_2 * u^2(t) \right] dt \quad (14)$$

Here ITAE criterion has been chosen because it is the only criterion whose selectivity is best out of all defined time domain performance criterion (ISE, ITSE and IAE) and also it has the best capacity to culminate the system output with small overshoot and oscillations. ISCO criterion is selected to minimize the effort of control input and perturbations havoc so that the saturation of the actuator can be avoided. In the above performance index optimization weighting functions $S_1$ and $S_2$ is chosen to be same so that error and control effort minimization have equal impact on the system's output. However, if a control designer wanted to achieve good overshoot and oscillations without taking care of control effort then he/she should put more weights to ITAE rather than to ISCO and vice versa.

The second performance criteria defined for the perturbed LQR system is the maximization of the singular value of the return ratio matrix at the input of the plants. This performance index can be formulated as following new lower bound of the singular values [5] - [6].

$$J_2 = \sum_{j=1}^{k_2} \sum_{i=1}^{k_1} \sqrt{\frac{\sigma_{min}^2(R^{0.5})}{\sigma_{max}^2(R^{0.5})} + \frac{\sigma_{min}^2(H(s))}{\sigma_{max}^2(R^{0.5})}} \quad (15)$$

Where $H(s) = Q^{0.5}(sI - A)^{-1}B$ and $k_I=min(m,n)$; *m* and *n* are the number of inputs and outputs of the MIMO system and $\sigma$ are the singular values of the corresponding defined matrices for the range of frequencies $\omega_1 \leq \omega_j \leq \omega_{k_2}$. The above formulation of the performance index will change for single input systems i.e. *R* matrix has only one element and hence only one scalar singular value can be generated. Hence, the formulation of $J_2$ will be done in the following equation for a single input.

$$J_2 = \sum_{j=1}^{k_2} \sum_{i=1}^{k_1} \sqrt{1 + \frac{\sigma_{min}^2(H(s))}{\sigma_{max}^2(R^{0.5})}} \quad (16)$$

The second performance index $J_2$ is the new bound for the return difference matrix at the plant's input and robust stability of the system can be ensured if norm of the $J_2$ will be greater than or equal to 1 as proved [6]. Also, asymptotic stability will be possible if the pair [$Q^{0.5}$, $Perturbed\,System$] is detectable.

The third formulation of performance index is the minimization of maximum singular values of the perturbation transfer function at the plant's output i.e. there should not be any closed loop right half plane zeros for $s \in Nyquist\,D\text{-}Contour$. The formulation is in the following equation.

$$J_3 = \sum_{j=1}^{k_2} \sum_{i=1}^{k_1} \sigma_{max} \left[ \frac{1-L(s)}{L(s)} \right] \qquad (17)$$

Where $k_1=min(m,n)$; $m$ and $n$ are the number of inputs and outputs of the MIMO system and $\sigma$ are the singular values of the corresponding defined matrices for the range of frequencies $\omega_1 \leq \omega_j \leq \omega_{k_2}$.

So, the vector-valued performance index can now be written as given in the equation (18).

$$\vec{J} = [J_1\ J_2\ J_3]$$

Now the equation (18) has been solved with PESA-II algorithm with population size taken as 200 and the number of generations is 250.

### 4. CASE STUDIES AND DISCUSSIONS

Two case studies have been presented in this paper to validate the proposed design of LQR systems. The system takes into consideration is shown by the equation (6) and another is shown by the equation (8). The commensurate order state-space models of the systems under consideration are represented by equations 7 and 9. The simulation results are computed in MATLAB 2016a. The weighting matrices Q and R considered in the performance index must satisfy the positive semi-definiteness and positive definiteness. When these conditions are fulfilled along with detectability of the pair [$Q^{0.5}$, $Perturbed\,System$] then matrix P has been computed from the Riccatti equation and must satisfy the positive definiteness condition. For the solution of LQR system the vector-valued performance measure has been solved with two objective functions taken at a time and finally, all the objective functions considered simultaneously. The trade-off has been made between the objective functions while solving for the Pareto optimal solutions.

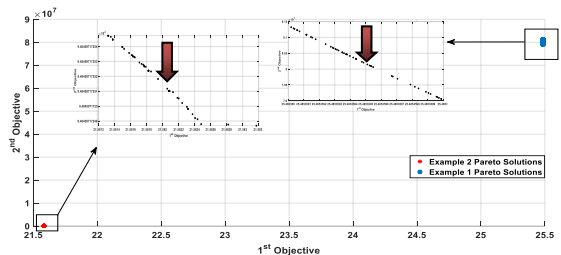
Figure 2. Pareto Front of the system for $J_1$ and $J_2$

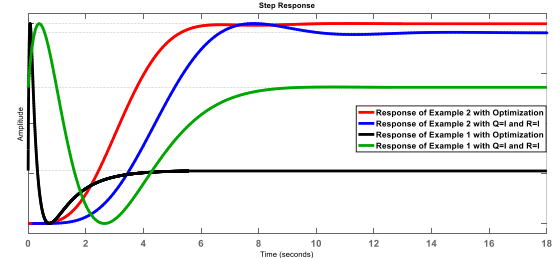
Figure 3. Step Responses of the best controller solution for $J_1$ and $J_2$

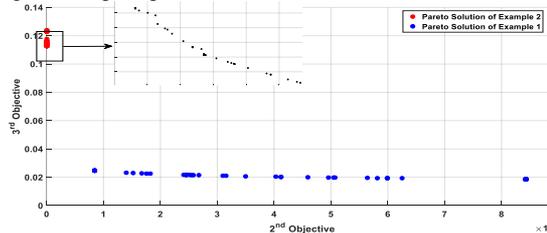
Figure 4. Pareto Front of the system for $J_2$ and $J_3$

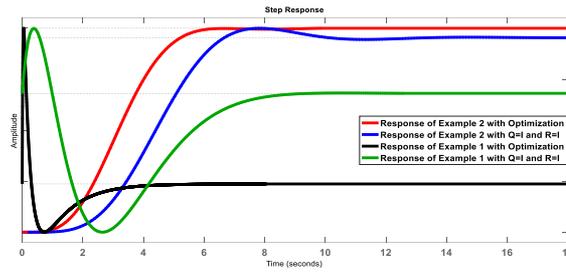

Figure 5. Step Responses of the best controller solution for $J_2$ and $J_3$

In the first attempt, J1 and J2 are considered for good time domain performance, optimal control energy expenses and robust stability of the system with minimum perturbations effect. The best controller is selected among the non-dominated solutions after trade-off between the objective functions. A good trade-off for both the objective functions is to select a controller from the area indicated by red colored arrow on the magnified Pareto solutions represented in figure 2. The time responses of the controllers are shown in figure 3. The time response plots show the comparison between the responses for the commensurate order systems (with the optimized value of $Q$ and $R$ and without optimization by selecting $Q=I$ and $R=I$). Settling time and tuned values of the design variables (weighting matrices) for the commensurate order systems represented by equations 7 and 9 are shown in the tables. Similarly, $J_2 - J_3$, $J_1 - J_3$ objectives have been solved in two dimensions by making trade-offs. Pareto fronts and the responses are shown in the figures 4, 5, 6, and 7 respectively. Finally, the entire objective functions ($J_1 - J_2 - J_3$) are solved simultaneously by making the trade-off between all the objective functions so that all the performance criteria defined in equations 14, 16 and 17 are satisfied in a symmetrical fashion. The Pareto front and response corresponding to the vector-valued objective function is shown in figure 8 and 9 respectively. In each of the examples considered we can see that the trade-off between the entire objective functions (i.e. $J_1 - J_2 - J_3$) gives a good result by making a compromise between extreme conditions. The weighting matrices variables value of the best controller solution after PESA-II optimization is shown in table I and table II with settling time which represents the speed of the response.

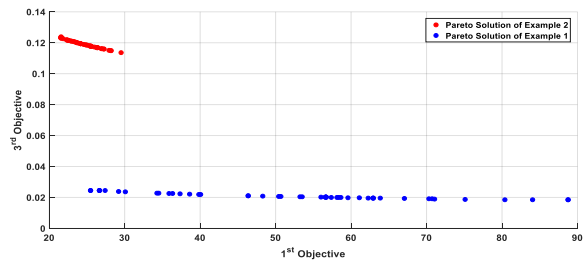

Figure 6. Pareto Front of the system for $J_1$ and $J_3$

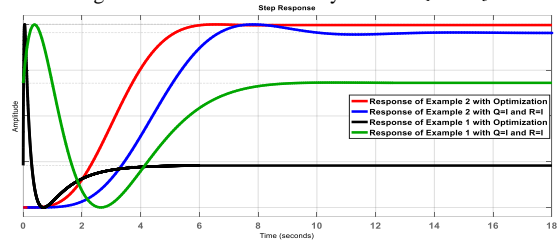

Figure 7. Step Responses of the best controller solution for $J_1$ and $J_3$

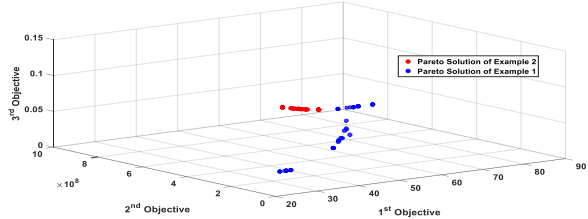

Figure 8. Pareto Front of the system for $J_1$, $J_2$ and $J_3$

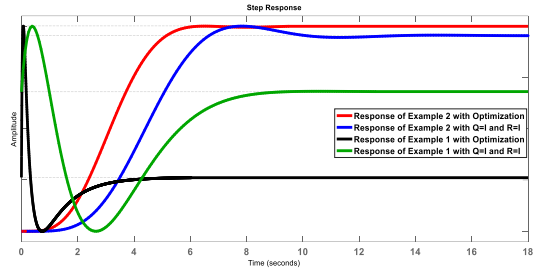

Figure 9. Step Responses of the best controller solution for $J_1$, $J_2$ and $J_3$

### B. Tables for the design variables obtained after simulation

**Table 1** [Example 1]

| Design Variables Table | Trade-off between performance Indices for the system shown in equation 7 with PESA-II | | | |
|---|---|---|---|---|
| | *J1-J2* | *J2-J3* | *J1-J3* | *J1-J2-J3* |
| Q1 | 0.5 | 0.531 | 0.500115 | 0.5 |
| Q2 | 0.0001 | 0.0071 | 0.0001006 | 0.01 |
| Q3 | 227.405 | 267.037 | 299.6071 | 269.2675 |
| R | 0.001 | 0.00033 | 0.000301 | 0.0001 |
| Settling time for best controller responses (sec.) | 3.72 | 3.57 | 3.38 | 3.71 |

a. For *Q=I* and *R=I* settling time is 8.16 sec.

**Table 2** [Example 2]

| Design Variables Table | Trade-off between performance Indices for the system shown in equation 9 with PESA-II | | | |
|---|---|---|---|---|
| | *J1-J2* | *J2-J3* | *J1-J3* | *J1-J2-J3* |
| Q1 | 1.001 | 1.0038 | 1.00106 | 1.004 |
| Q2 | 1.00056 | 1.00067 | 1.000571 | 1.0007 |
| Q3 | 0.02999 | 0.02665 | 0.0292 | 0.03 |
| Q4 | 3.5011 | 4.0000 | 3.5989 | 3.5 |
| Q5 | 5.7959 | 5.71401 | 5.9993 | 5.898 |
| Q6 | 4.0000 | 4.09789 | 4.2492 | 4.25 |
| R | 0.0020 | 0.0020 | 0.001431 | 0.0020 |
| Settling time for best controller responses (sec.) | 3.72 | 5.63 | 5.58 | 5.55 |

a. For *Q=I* and *R=I* settling time is 9.22 sec.

## 5. CONCLUSION

The LQ Regulator design for fractional order state-space systems has been proposed in this paper with the optimal selection of design parameters Q and R. The optimal selection of weighting matrices Q and R have been done through PESA-II optimization algorithm. The performance indices minimized for the optimal selections of design parameters are the weighted sum of ITAE and ISCO to avoid high oscillations and overshoots. Robust stability is achieved through the minimization of singular values of return ratio matrix at the plant's input, and the maximum singular values of the perturbation transfer function at the plant's output are minimized to avoid right half zeros in the system. Thus, the proposed design method of LQ Regulator preserves its state trajectories optimality, and also lowers the closed-loop error indexes in the presence of perturbations. The proposed techniques and simulation results show that it works for special classes of unstable fractional order state-space systems with uncertainties.